# Resonant x-ray scattering in 3d-transition-metal oxides: Anisotropy and charge orderings

G. Subías[1], J. García[1], J. Blasco[1], J. Herrero-Martín[2] and M. C. Sánchez[1]
[1] Instituto de Ciencia de Materiales de Aragón, Departamento de Física de la Materia Condensada, CSIC-Universidad de Zaragoza, 50009-Zaragoza, Spain
[2] European Synchrotron Radiation Facility, 38043-Grenoble-Cedex, France

E-mail: gloria@unizar.es

**Abstract**. The structural, magnetic and electronic properties of transition metal oxides reflect in atomic charge, spin and orbital degrees of freedom. Resonant x-ray scattering (RXS) allows us to perform an accurate investigation of all these electronic degrees. RXS combines high-Q resolution x-ray diffraction with the properties of the resonance providing information similar to that obtained by atomic spectroscopy (element selectivity and a large enhancement of scattering amplitude for this particular element and sensitivity to the symmetry of the electronic levels through the multipole electric transitions). Since electronic states are coupled to the local symmetry, RXS reveals the occurrence of symmetry breaking effects such as lattice distortions, onset of electronic orbital ordering or ordering of electronic charge distributions. We shall discuss the strength of RXS at the K absorption edge of 3d transition-metal oxides by describing various applications in the observation of local anisotropy and charge disproportionation. Examples of these resonant effects are (I) charge ordering transitions in manganites, $Fe_3O_4$ and ferrites and (II) forbidden reflections and anisotropy in $Mn^{3+}$ perovskites, spinel ferrites and cobalt oxides. In all the studied cases, the electronic (charge and/or anisotropy) orderings are determined by the structural distortions.

## 1. Introduction

The discovery of new electronic properties, including high-$T_C$ superconductivity, colossal magnetoresistance and multiferroic modifications, in transition-metal oxides has fuelled a resurgence of interest in atomic charge, spin and orbital degrees of freedom in systems of highly correlated electrons. X-ray spectroscopy techniques are highly sensitive to these electron degrees of freedom. However, they probe short-distance correlations only. In contrast, x-ray diffraction reveal long-range ordered static correlations. Resonant x-ray scattering (RXS) combines absorption and diffraction as they have in common the x-ray atomic scattering factor (ASF), $f$, which is usually written as: $f = f_0 + f' + if''$ [1]. It contains an energy independent part, $f_0$, corresponding to the classical Thomson scattering and two energy-dependent terms, $f'$ and $f''$, also known as the anomalous ASF. RXS occurs when the x-ray energy is tuned near the absorption edge of an atom in the crystal. In this case, the anomalous ASF strongly depends on the photon energy, which manifests in marked variations of the scattered intensity. This dependence of the scattered intensity appears in any Bragg reflection when crossing the absorption edge of a constituent atom.

The study of the energy-dependent modulation of the diffraction intensity of intense Bragg peaks is the scope of the diffraction anomalous fine structure technique (DAFS) [2,3]. This technique allows the determination of the local structural information around the anomalous atom that is chemical and valence specific similar to that of Extended X-ray Absorption Fine

Structure (EXAFS). The advantage of DAFS is that it is spatially and site-selective. Our interest here is focused on the study of either weak-allowed or forbidden reflections that appear in a phase transition. In the first case, the anomalous scattering contribution is comparable to the Thomson one and the peculiar characteristics of RXS can be studied. The intensity of weak superlattice reflections can be due to either a structural modulation, contributing to the structure factor as a Thomson term or because the anomalous ASF of atoms, now in different crystallographic sites, differ in some energy range. Normally, these differences are larger for photon energies close to an absorption edge, showing an enhancement (or strong decrease) of the scattered intensity just around the absorption threshold. In the case of symmetry forbidden reflections, only the resonant term is present. Since RXS involves virtual transitions of core electrons into empty states above the Fermi level, the excited electron is sensitive to any anisotropy around the anomalous atom so that the anomalous ASF has a tensorial character. Thus, a reflection can be observed on resonance if any of the components of the structure factor tensor is different from zero. In the case of weak superlattice reflections, the resonant atoms occupy different crystallographic sites and have different local structures. Thus, the anomalous ASF are different at energies close to the absorption edge and the scattered intensity shows a resonance reflecting the differences of the ASF between these non-equivalent atoms. RXS intensity is then observed coming from the difference between the diagonal terms of the ASF. In this case, there is a Thomson contribution and the analysis of the spectral shape must include it. On the other hand, resonances observed in forbidden reflections are related to atoms that occupy equivalent crystallographic sites. Symmetry elements with translation components (screw axes and glide planes) of the crystal space group transform an atom into another equivalent in the lattice, but with a differently oriented atomic surrounding. This makes that some of the off-diagonal terms change sign after these symmetry operations, giving rise to structure factor tensors that contain non-vanishing off-diagonal terms. Templeton & Templeton [4] first noticed such reflections that show a local polarization anisotropy of the x-ray susceptibility and they are now known as ATS reflections. We note that ATS reflections have strong polarization and azimuth dependences.

The vector potential of the electromagnetic field in the matter-radiation interaction term can be developed in a multipolar expansion in such a way that the symmetry of the excited levels can be chosen. Thus, we can speak on dipolar-dipolar transitions, dipolar-quadrupolar transitions, etc. Consequently, resonant methods are also sensitive to the symmetry of the electronic shells, which compose the intermediate states. For electric dipole-dipole (E1E1) transitions the wave vectors $k_i$ and $k_f$ of the incident and scattered x-rays do not enter the scattering amplitude. The strength of the x-ray resonances associated with electric quadrupole-quadrupole (E2E2) and electric dipole-quadrupole transitions (E1E2) are less intense than for E1E1 process but must content the $k_i$ and $k_f$ dependence. A complete polarization and azimuthal analysis of the RXS experiment is needed to assign the multipolar origin of the RXS signal.

In this contribution, after a recall of the RXS amplitude, we will discuss several examples that correlate with the charge and orbital ordering concepts. We restrict to the metal K edges and mainly to the dipolar-dipolar channel, which describes most of the phenomenology. Two types of reflections are observed. The first one originates from the different anomalous scattering factors of non-equivalent atoms in the crystal. The occurrence of this type of resonance is correlated to the energy shift of the absorption edge (chemical shift) and it has been considered as a proof of charge ordering (CO). The second one arises from the anisotropy of the anomalous scattering factor of an atom at crystallographic equivalent sites. These ATS reflections have been assigned to orbital ordering (OO). However, the first type of reflections can also exhibit anisotropic behaviour. In all the studied cases, we conclude two main results: (i) the charge segregation is much smaller than one electron and consequently, it would be better described as a charge modulation and (ii) the anisotropy is present in the p-empty density of states and it seems not to be correlated with a real d-orbital ordering.

## 2. Resonant x-ray scattering tensor
In RXS, the global process of photon absorption, virtual photoelectron excitation and photon re-emission, is coherent through the crystal, giving rise to the usual Bragg diffraction condition

$$F = \sum_j e^{i\vec{Q}\cdot\vec{R}_j}(f_{0j} + f'_j + if''_j) \qquad (1)$$

where $\vec{R}_j$ is the position of the $j$-th scattering atom in the unit cell, $\vec{Q} = \vec{k}^f - \vec{k}^i$ is the scattering vector ($\vec{k}^i$ and $\vec{k}^f$ are the wave vectors of the incident and scattered beams) and $f_{0j}$ is the Thomson scattering part of the atomic scattering factor. The resonant part, $f' + if''$, is given by the expression [5]

$$f' + if'' = \frac{m_e}{\hbar^2}\frac{1}{\hbar\omega}\sum_n \frac{(E_n - E_g)^3 \langle\psi_g|\hat{O}^{f*}|\psi_n\rangle\langle\psi_n|\hat{O}^i|\psi_g\rangle}{\hbar\omega - (E_n - E_g) - i\frac{\Gamma_n}{2}} \qquad (2)$$

In this expression, $\hbar\omega$ is the photon energy, $m_e$ is the electron mass, $\psi_g$ describes the initial and final electronic state with energy $E_g$ and $E_n$ and $\Gamma_n$ are the energy and inverse lifetime of the intermediate excited states $\psi_n$. The interaction of the electromagnetic radiation with matter is expressed by the operators $\hat{O}^i$ and $\hat{O}^{f*}$. By multipole expansion of these operators up to the electric quadrupole term, we have [6]:

$$\hat{O}^{i(f)} = \vec{\varepsilon}^{i(f)} \cdot \vec{r}(1 - \frac{1}{2}i\vec{k}^{i(f)} \cdot \vec{r}) \qquad (3)$$

Here $\vec{r}$ is the electron position measured from the absorbing atom and $\vec{\varepsilon}^{i(f)}$ is the polarization of the incident(scattered) beam. Correspondingly, we get that there are three contributions to the resonant scattering factor: dipole-dipole (*dd*), dipole-quadrupole (*dq*) and quadrupole-quadrupole (*qq*).

Using the Cartesian reference coordinate system defined in figure 1 for the photon polarization and wave vector, we can develop the scalar product of equation (3) and the resonant scattering amplitude of equation (2) can be written in the form:

$$f' + if'' = \frac{m_e}{\hbar^2}\frac{1}{\hbar\omega}\sum_n \frac{(E_n - E_g)^3}{\hbar\omega - (E_n - E_g) - i\frac{\Gamma_n}{2}}$$
$$\left[\sum_{\alpha,\beta}\varepsilon_\alpha^{f*}\varepsilon_\beta^i D_{\alpha\beta} - \frac{i}{2}\sum_{\alpha\beta\gamma}\varepsilon_\alpha^{f*}\varepsilon_\beta^i(k_\gamma^i I_{\alpha\beta\gamma} - k_\gamma^f I^*_{\beta\alpha\gamma}) + \frac{1}{4}\sum_{\alpha\beta\gamma\delta}\varepsilon_\alpha^{f*}\varepsilon_\beta^i k_\gamma^f k_\delta^i Q_{\alpha\beta\gamma\delta}\right] \qquad (4)$$

where $\alpha$, $\beta$, $\gamma$, $\delta$ are indexes that vary independently over the three Cartesian directions $x$, $y$, $z$, and the transition matrix elements $D_{\alpha\beta}$, $I_{\alpha\beta\gamma}$ and $Q_{\alpha\beta\gamma\delta}$ associated to *dd*, *dq* and *qq* contributions are characterized by the following Cartesian tensors of second, third and fourth rank, respectively:

$$D_{\alpha\beta} = \sum_n \langle\psi_g|r_\alpha|\psi_n\rangle\langle\psi_n|r_\beta|\psi_g\rangle$$
$$I_{\alpha\beta\gamma} = \sum_n \langle\psi_g|r_\alpha|\psi_n\rangle\langle\psi_n|r_\beta r_\gamma|\psi_g\rangle \qquad (5)$$
$$Q_{\alpha\beta\gamma\delta} = \sum_n \langle\psi_g|r_\alpha r_\beta|\psi_n\rangle\langle\psi_n|r_\gamma r_\delta|\psi_g\rangle$$

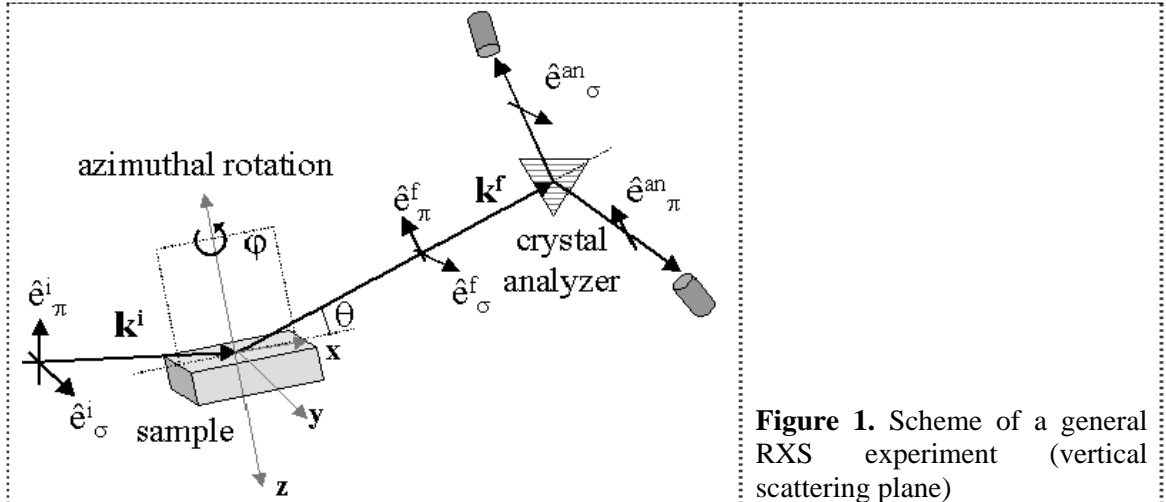

**Figure 1.** Scheme of a general RXS experiment (vertical scattering plane)

It is important to determine the symmetry properties of the *D*, *I* and *Q* tensors [7] that depend on two factors: (i) the transformation properties of a space group itself and (ii) the local symmetry of the resonant atoms position. Referring to non-magnetic samples, the structure factor tensor is a second rank symmetric tensor (parity-even) with six independent components in the *dd* approximation. This number reduces upon taking into account the local site symmetry of the atom. The *qq* term is also symmetrical under the inversion symmetry, whereas the *dq* contribution is only antisymmetrical. As a result, if an atom sits in an inversion centre, the *I* tensor must be zero and the dipole-quadrupole transition is only allowed for atoms breaking the inversion symmetry.

In the following we shall describe some significant RXS experiments following the two types of resonant reflections, weak-allowed and forbidden. In the first case, we shall relate them to the charge modulation, intimately joined to the transition metal-ligand bond distances and in the second case we shall relate them to the anisotropy of the geometrical local structure around the resonant atoms. Since the charge is a time-reversal invariant quantity, we shall deal either with pure electric *dd* and *qq* transitions that are even under parity or with parity-odd electric *dq* transitions. Time-reversal odd events that are related to the magnetic properties of the system will be not considered.

## 3. Resonant effects due to different crystallographic sites (charge) ordering

Understanding the charge state in the high and low temperature phases of mixed valence transition-metal oxides is of fundamental interest in the context of metal-insulator transitions that are assumed to be driven by CO. The sensitivity of RXS to the CO relies on the fact the energy values of the absorption edge for the two different valence states of the transition-metal atom are slightly different (known as chemical shift). If long-range order of these valence states exists, superlattice reflections due to the contrast between the atomic scattering factors of the two valence states will exhibit a resonance enhancement. The point is that CO is intimately correlated with the associated crystal distortions coming from the structural transitions that accompanied the metal-insulator ones. Thus, a question arises whether the electronic CO is the cause or the effect of the lattice distortions.

The concept of CO in solids was first applied by E. J. W. Verwey to the metal to insulator transition that occurs in magnetite ($Fe_3O_4$) at $T_v \sim 120$ K, now known as the Verwey transition [8]. Above $T_v$, $Fe_3O_4$ has the inverse spinel cubic $AB_2O_4$ structure, where A and B are the tetrahedral and octahedral Fe sites, respectively. Verwey originally proposed that the hopping of valence electrons on the octahedral B-site sublattice is responsible for metallic conductivity. In the insulating phase, spatial localization of the valence electrons on these B-sites gives rise to an ordered pattern of $Fe^{3+}$ and $Fe^{2+}$ ions in successive [001] planes (cubic notation). The B-site sublattice, shown in the inset of figure 2, can be regarded as a diamond lattice of tetrahedra of nearest-neighbour Fe atoms sharing alternate corners. This simple model was implying the observation of (0,*k*,*l*) reflections with *k*+*l*=4n+2 in the low temperature phase. These reflections are forbidden above $T_v$ because of the diamond glide plane of the spinel structure. The first RXS experiments in magnetite showed that (002) and (006) reflections belonging to this type of cubic forbidden reflections originates from the local structural anisotropy of the Fe atoms at the B-sites that have a trigonal point symmetry ($\bar{3}m$) [9-11]. These works discarded the Verwey's CO model but they did not guarantee the lack of CO with other periodicities of the cubic unit cell.

In order to investigate other possible CO periodicities, we need to start from the low temperature crystallographic structure. The symmetry lowering $Fd\bar{3}m \rightarrow Cc$ generates 8 and 16 non-equivalent Fe sites at tetrahedral and octahedral positions, respectively; each one can have its own local atomic charge. However, the exact structure is not yet perfectly known [12-14]. A good approach to the real structure consists of a $P2/c$ cell with lattice parameters $\approx a_c/\sqrt{2} \times a_c/\sqrt{2} \times 2a_c$, $a_c$ being the cubic cell parameter [12,14]. Complexity is greatly reduced because there are only 6 non-equivalent Fe atoms, two in tetrahedral sites ($A_1$ and $A_2$) and four in octahedral sites ($B_1$, $B_2$, $B_3$ and $B_4$). It can be noticed that atomic displacements in this low temperature structure result in two main types of superlattice reflections, which are indexed in the cubic notation: (I) ($h,k,l$) reflections such that $h+k$=even resulting form the loss of the $fcc$ translation that give rise to charge modulations with wave vector $\vec{q} = (0,0,1)c$ and (II) half-integer ($h,k,l+1/2$) reflections arising from the doubling of the cell along the $c$ axis that corresponds to charge modulation with $\vec{q} = (0,0,1/2)c$. We examine now the limit for the possible charge segregation over the octahedral atoms along the $c$ axis given by the sensitivity of the RXS technique in a highly stoichiometric single crystal of $Fe_3O_4$ ($T_V$=123.5 K) [15].

Figure 2 (left panel) shows the energy dependence of the intensities for some characteristic Bragg and forbidden reflections at the Fe K-edge and at 60 K compared to the fluorescence spectrum. Experimental data have been corrected for absorption.

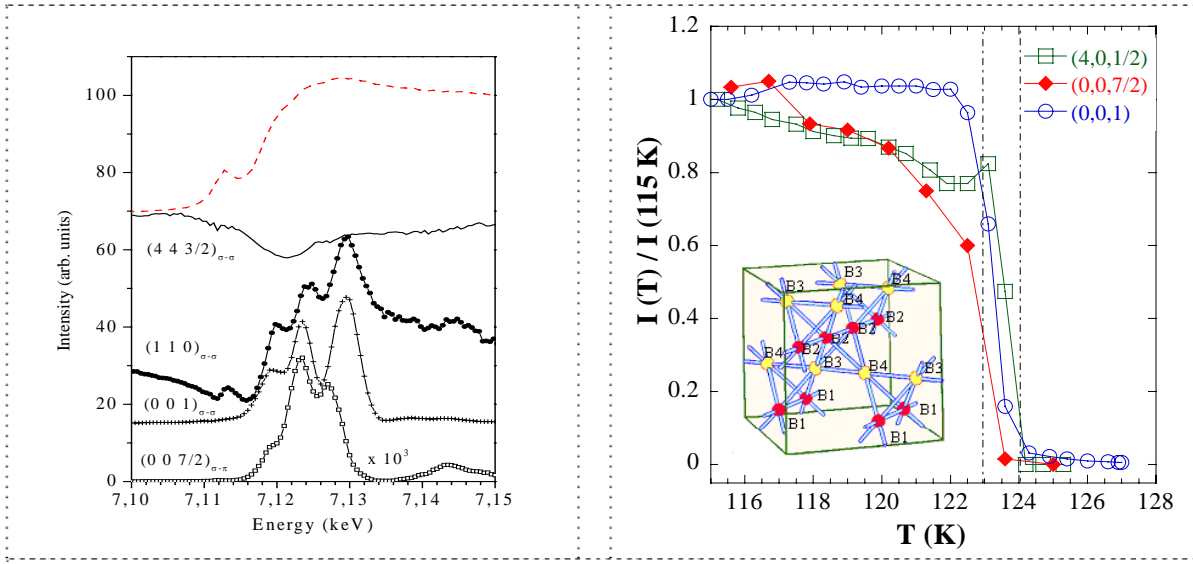

**Figure 2.** The left panel shows some of the experimental RXS spectra around the Fe K-edge in $Fe_3O_4$, corrected for absorption. The right panel shows the temperature dependence of the integrated intensities, on and off-resonance, normalized to the low-temperature value.

(a) Energy scans of the permitted (0,0,1) and (1,1,0) reflections show three resonant peaks, a first one at the Fe K threshold (7118 eV) and the other two around 7124-7129 eV, which corresponds to the white line in the fluorescence spectrum. The observed strong resonant effect is a consequence of electronic and structural differences among the Fe atoms at B1 and B2 octahedral sites. This can be parameterised in terms of valence as a charge segregation δ=0.23$e$. This result confirms the lack of ionic CO in terms of $Fe^{2+}$ and $Fe^{3+}$ ions, in agreement with previous RXS [16-18] and synchrotron powder diffraction [14] studies.

(b) Resonant intensity is only observed in the σ−π' channel for the (0,0,7/2) reflection, indicating that this is a forbidden reflection in the low temperature phase. The energy scan shows a three-peak structured resonance nearly at the same energies as the (0,0,1) and (1,1,0) reflections. In this case, the electronic anisotropy comes from interference among equivalent crystallographic sites. Six different sites are present for the Fe atoms so up to six different terms could contribute to the resonant signal [19]. The superlattice (4,4,3/2) corresponds to the same periodicity along c as the forbidden (0,0,*l*/2) reflections. However, it displays hardly any resonant effect opposite to what is expected to occur at those very weak reflections. Therefore, we can conclude that no charge segregation exists with (0,0,1/2) periodicity and the ATS (0,0,*l*/2) reflections have its origin in the loss of octahedral and tetrahedral symmetry originated by the structural phase transition.

In order to establish the correlation between the lattice distortion and the charge segregation and anisotropy orderings, we have measured the temperature dependence of the intensity of the following superlattice reflections: (4,0,1/2) off-resonance (E=7.1 keV) and (0,0,1) and (0,0,7/2) on resonance (E=7.125 keV), which is reported in figure 2 (right panel). The resonant and non-resonant signals simultaneously disappear at $T_v$ (±0.5 K) and the intensity of all these reflections is zero at temperatures above 125 K. This result shows that RXS in $Fe_3O_4$ comes from the ordering of local distortions at the structural transition, which leads to an ordered formal charge segregation and electronic anisotropy at the Fe atoms [15].

We will comment now on the half-doped manganites such as $Nd_{0.5}Sr_{0.5}MnO_3$ and $Bi_{0.5}Sr_{0.5}MnO_3$. It was proposed the ordering of an alternating pattern of $Mn^{3+}$ and $Mn^{4+}$ ions was predicted leading to the onset of superlattice reflections doubling the *b* axis of the orthorhombic *Pbnm* (*Ibmm*) cell [20]. The observed modulation wave vectors are (0,*k*,0) and (0,*k*/2,0) with *k* odd for the ordering of charge and orbital degrees of freedom, respectively. The occurrence of orbital order (OO) is also predicted independently from the CO because $Mn^{3+}$ are Jahn-Teller distorted ions. We have measured the energy dependence of the intensity at the Mn K edge at Q=(0,3,0) and Q=(0,5/2,0) for both, Nd:Sr [21] and Bi:Sr [22] single crystals. Figure 3 summarizes the results obtained.

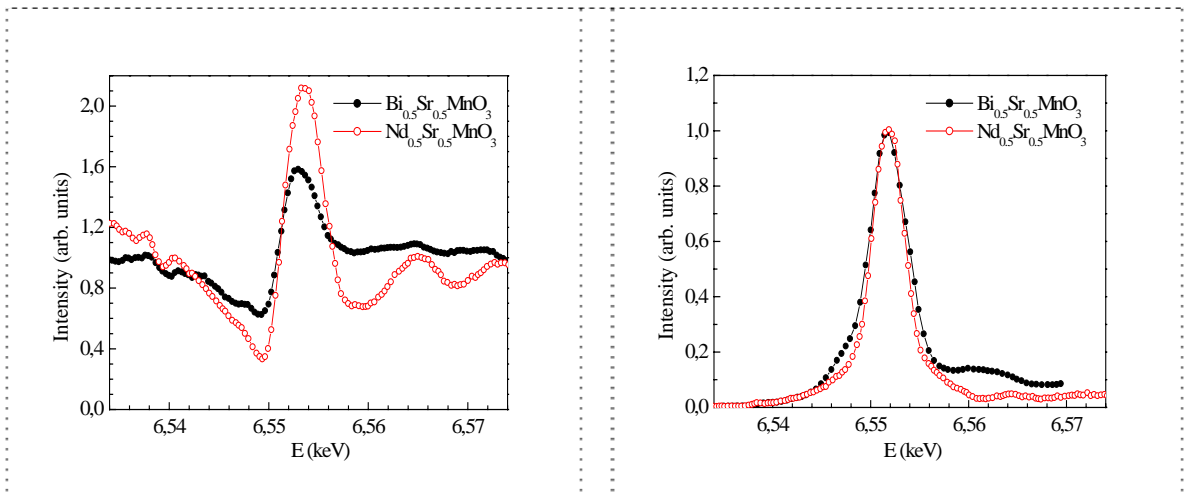

**Figure 3.** The left panel shows the normalized intensity for the non-rotated polarization channel at the (0,3,0) peak associated to CO whereas the right panel shows the rotated polarization channel at the (0,5/2,0) peak associated to OO in $Nd_{0.5}Sr_{0.5}MnO_3$ and $Bi_{0.5}Sr_{0.5}MnO_3$.

Non-resonant intensity can be observed away from the K-edge of Mn and a broad resonance is found at the absorption edge at (0,3,0) reflections in the non-rotated (σ-σ') channel. On the other hand, a strong Gaussian-shaped resonance is observed at energies close to the K-edge of Mn at (0,5/2,0) reflections only in the rotated (σ-π') channel, which identifies these half-integer reflections as structurally forbidden ones. The variation with azimuth of these resonances shows a characteristic oscillation with π periodicity. We note that in this case, the two kinds of behaviour mixed since the marked different anisotropy of the two types of atoms. The resonant scattering at either (0,$k$,0) or (0,$k$/2,0) reflections does not depend on the $k$ wave vector and it arises from E1 (1s→ np) transition. The temperature dependence of the resonant intensities shows that the two types of reflections disappear at the metal-insulator phase transition.

The complete analysis of the energy line-shape and the azimuth and polarization dependence of the resonant intensities is carried out using a semi-empirical structural model [23]. The checkerboard arrangement of two types of crystal Mn sites gives an excellent agreement with the experimental data. These two sites differ in their local geometric structure: one site is *anisotropic* (tetragonal distorted oxygen octahedron) and the other site is *isotropic* (nearly undistorted oxygen octahedron). Estimates of the valence modulation between the two Mn atoms are slightly variable depending on the manganite but all fall below the ideal charge segregation of ±0.5$e$. As shown in [21] and [22], ±0.08 for $Nd_{0.5}Sr_{0.5}MnO_3$ and ±0.07 for $Bi_{0.5}Sr_{0.5}MnO_3$, respectively.

CO for x different from 0.5 is in general not well defined. We have also explored further the $Bi_{1-x}Sr_xMnO_3$ system toward the Mn3+-rich side, that is for x=0.37 [24]. Superlattice reflections of (0,$k$,0) and (0,$k$/2,0) types with $k$ odd were observed at the Mn K-edge. This is to be compared with the observation of the K-edge resonances in $Bi_{0.5}Sr_{0.5}MnO_3$. Figure 4 shows an example of the resonant enhancement as observed at (0,3,0) and (0,7/2,0) reflections in $Bi_{0.63}Sr_{0.37}MnO_3$.

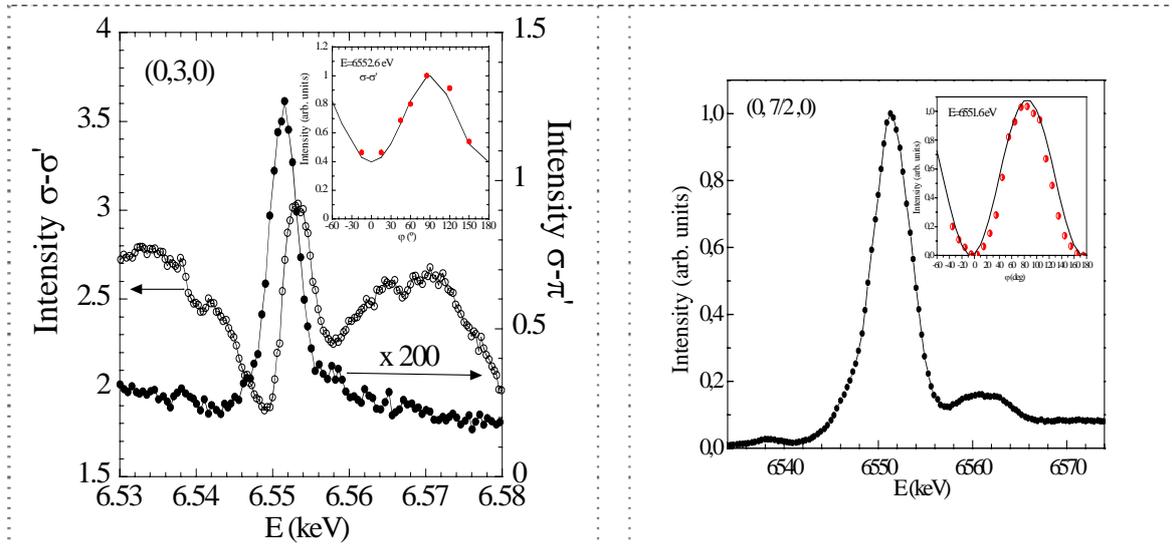

**Figure 4.** RXS results in $Bi_{0.63}Sr_{0.37}MnO_3$ at the Mn K-edge. (a) Polarization analysis of the (0,3,0) reflection as a function of photon energy. (b) energy dependence of the forbidden (0,7/2,0) reflection for the σ-π' channel. The insets show the respective azimuthal angle dependence on resonance.

The energy dependence of the intensity for both, weak-allowed (0,$k$,0) and forbidden (0,$k$/2,0) reflections is identical to that observed in the half-doped bismuth manganite (figure 3). The dependencies of the resonant intensities with azimuthal angle reveals identical twofold symmetry too. We note that the σ-σ' intensity of the (0,3,0) peak approaches the non-resonant intensity at the minimum opposite to the constant evolution expected for a pure CO reflection. These results indicate that the checkerboard ordering of two types of Mn atoms in terms of the local structure in the *ab* plane in a ratio 1:1 is strongly stable and extends to doping concentrations x<0.5. Intermediate valence states lower than +3.5 are deduced for $Bi_{0.63}Sr_{0.37}MnO_3$, where the charge disproportion is found to be ±0.07.

The last example is provided by $La_{0.33}Sr_{0.67}FeO_3$, which shows a charge modulation that is commensurate with the carrier concentration (n=1-x) and corresponds to a wave vector q=(2π/$a_p$) (1/3,1/3,1/3), being $a_p$ the primitive cubic lattice parameter. Ordered layers of $Fe^{3+}$ and $Fe^{5+}$ ions in a sequence of …$Fe^{3+}Fe^{3+}Fe^{5+}$… along the cubic [111] direction were originally proposed to explain the metal-paramagnetic to insulator-antiferromagnetic transition at 200 K [25]. We have investigated this three-fold CO using the RXS technique [26]. Superlattice (h/3,h/3,h/3) reflections in the pseudocubic cell notation were observed in the non-rotated σ-σ' polarization channel for h=2, 4 and 5. These reflections are forbidden in both cubic (*Pm-3m*) and rombohedral (*R-3c*) symmetries. However, they exhibit a resonant behavior at energies close to the Fe K edge on top of a non-resonant signal, as shown in figure 5.

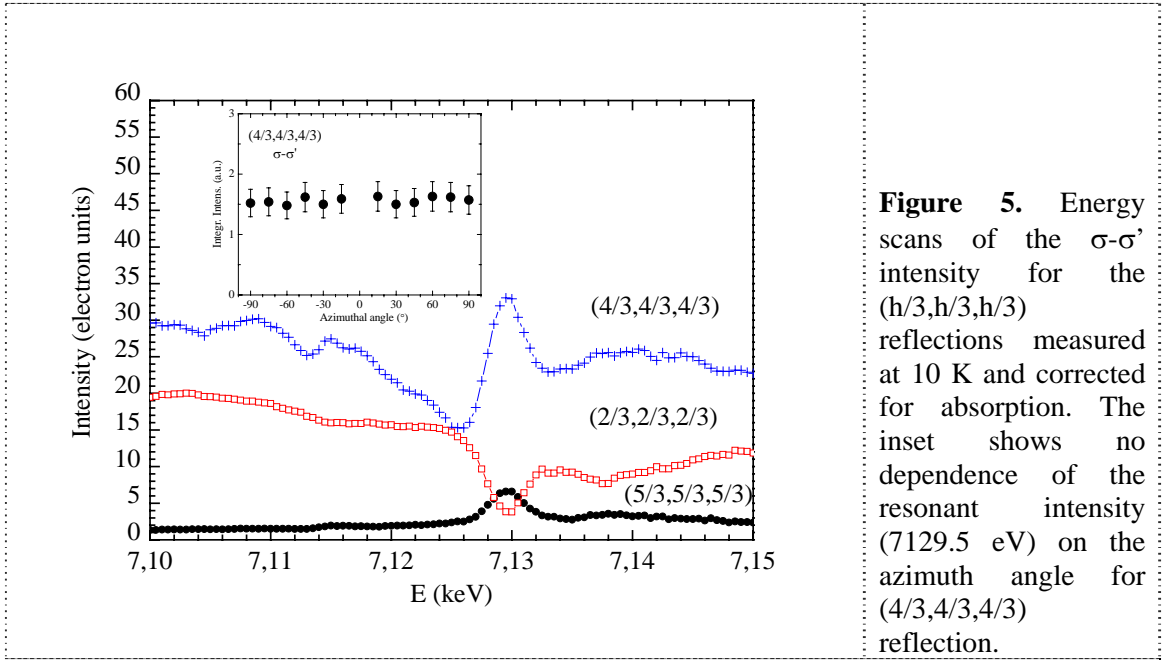

**Figure 5.** Energy scans of the σ-σ' intensity for the (h/3,h/3,h/3) reflections measured at 10 K and corrected for absorption. The inset shows no dependence of the resonant intensity (7129.5 eV) on the azimuth angle for (4/3,4/3,4/3) reflection.

The non-resonant intensities are explained due to a periodic structural modulation of both Fe and Sr(La)$O_3$ atomic planes along the cubic [111] direction. To account for the strong variation in the Thomson scattering among the different satellite reflections, the Fe and Sr(La)$O_3$ atomic planes must move in opposite senses, keeping the inversion symmetry of the supercell. These shifts differentiate two crystallographic sites for the Fe atoms and produce an ordered sequence of two compressed and one expanded Fe$O_6$ octahedra. The compressed octahedron is asymmetric with three short and three long Fe-O bonds whereas the expanded one is regular.

Concerning the local charges of the two types of Fe cations, the chemical shift between the expanded and compressed Fe atoms was found to be 0.7±0.1 eV. This is to be compared to the energy shift of 1.26 eV between $Fe^{3+}$ and $Fe^{4+}$ [27]. Assuming a linear relationship between energy shift of the K absorption edge and local charge, we determined that the amount of charge segregation is 0.6*e*. Since the resonant intensity is constant as a function of the azimuth angle over a range of 180 °, as shown in the inset of figure 5, the Fe atoms have not a local anisotropy in this case. This result indicates the presence of a pure charge density wave ordering in the mixed-valent perovskite $La_{0.33}Sr_{0.67}FeO_3$.

**4. Resonant effects due to anisotropy (orbital) ordering**
The direct observation of OO is a difficult task because it is accompanied by other effects such as lattice distortions and charge segregations. The study of the RXS and the anisotropic character of the x-ray scattering were developed in crystallography more than 20 years ago [4]. In particular, Dmitrienko [28,29] developed a general theoretical treatment of the "forbidden" ATS reflections, deriving new extinction rules valid near the absorption edge. In the microscopic point of view, these ATS reflections are due to the presence of the absorbing atom in an anisotropic chemical environment, which brings about the orientation of unoccupied electronic levels.

The experimental application of RXS to the study of OO started in 1998, when Murakami and co-workers [30] investigated the CO and OO in $La_{0.5}Sr_{1.5}MnO_4$ at the Mn K-edge. It was concluded the successful observation of OO of the $e_g$ electrons ($(3z^2-r^2)$- and $(x^2-y^2)$-type orbital) on $Mn^{3+}$ sites by detecting the ($2n+1/4,2n+1/4,0$) forbidden reflections. The angular dependence around the scattering vector of the intensity (azimuthal dependence) confirms that they result from the anisotropic character of the anomalous part of the atomic scattering factor. However, it was later demonstrated that the Jahn-Teller distortion of the oxygen octahedra surrounding the Mn atoms is sufficient to reproduce the experimental results at the K edge without invoking any contribution of the 3d-OO [31]. After this first observation, RXS is widely used to study the orbital degree of freedom in several manganites [32-36] and other transition-metal oxides [37-40]. In particular, the following experiments provide key information for the mechanism of RXS and OO.

(1) The parent compound of colossal magnetoresistive manganites, $LaMnO_3$, was expected to show an alternating ordering of $(3x^2-r^2)$ and $(3y^2-r^2)$ orbitals in the *ab* plane below 780 K due to the greatly distorted $MnO_6$ octahedron and the A-type antiferromagnetic order [41]. The observation of resonance in the σ-π' scattered intensity of the (3,0,0) forbidden reflection at the Mn K-edge was initially interpreted as a direct probe of this OO [32]. Moreover, the increase of the resonant signal with decreasing temperature as $T_N$ was approached from above was interpreted as a direct correlation of the magnetic order with OO.

Recently, we have revisited the RXS at the Mn K-edge of $LaMnO_3$, both experimentally and theoretically [36]. We have observed two independent forbidden reflections, (0,3,0) [or, equivalently (3,0,0)] and (0,0,3), which are related to two different nonzero off-diagonal elements of the second-rank atomic scattering factor tensor. The different energy dependence of the RXS spectra for the two types of forbidden reflections has been explained within the multiple scattering theory in terms of long-range ordered structural distortions around Mn atoms using a cluster that includes up to 63 atoms beyond the first oxygen neighbors [42], as demonstrated in figure 6(a). Since no change in either the intensity or its azimuthal dependence was observed when crossing $T_N$ at 140 K (see figure 6(b)), these forbidden reflections are ascribed as ATS reflections in agreement with a structural origin and opposite to the OO interpretation.

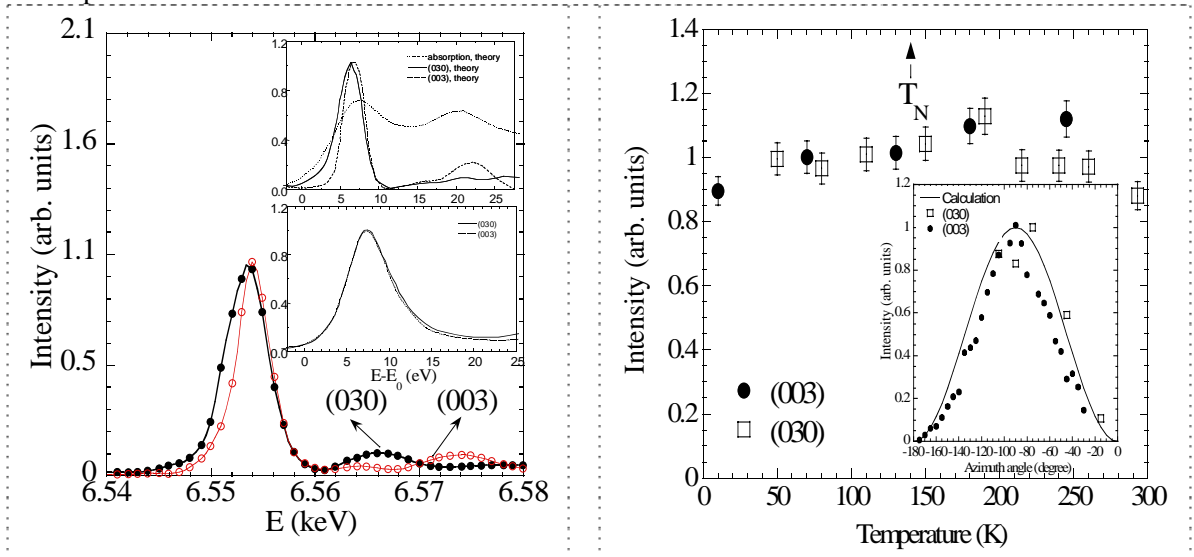

**Figure 6.** (a) Experimental (circles) RXS of the (0,3,0) and (0,0,3) forbidden reflections in $LaMnO_3$. The inset shows the MXAN calculations done using the 63-atoms cluster (upper panel) and the 6-atoms cluster (lower panel). (b) Temperature dependence of the integrated intensity of the (0,3,0) and (0,0,3) reflections on resonance, normalized for comparison. The inset shows the respective azimuthal angle dependence on resonance.

(2) Another example of the excitement of forbidden reflections on a cubic crystal structure that deserves discussion are the (0,0,l) (l=4n+2) reflections in $Fe_3O_4$. Once the Verwey model of

CO was discarded, these reflections offer the possibility to study RXS excited by a *dq*-transition which is only allowed on the tetrahedral site whereas the octahedral site allows, on the contrary, only for *dd* and *qq* transitions. Two distinct resonant lines are observed in the RXS spectra of (0,0,2) and (0,0,6) forbidden reflections in $Fe_3O_4$ at the pre-edge and at the main edge energies of the Fe K-edge [10]. The comparison of the energy and azimuthal dependence of these reflections at the Fe and Co K edges in $Fe_3O_4$, $CoFe_2O_4$ and $MnFe_2O_4$ spinels is reported in figure 7 [38]. No pre-edge resonance is observed either at the Fe K-edge in $MnFe_2O_4$ or at the Co K-edge in $CoFe_2O_4$, whereas the main-edge resonance is observed at the Fe K-edge in both, $CoFe_2O_4$ and $MnFe_2O_4$, and at the Co K-edge in $CoFe_2O_4$.

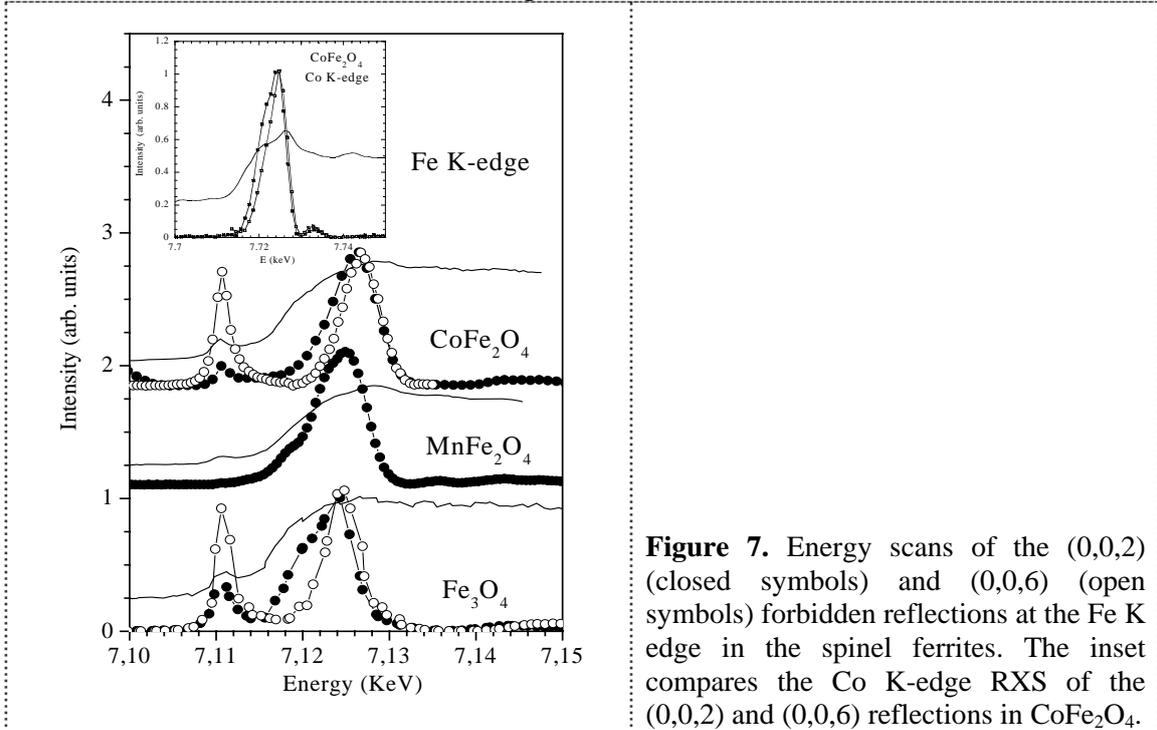

**Figure 7.** Energy scans of the (0,0,2) (closed symbols) and (0,0,6) (open symbols) forbidden reflections at the Fe K edge in the spinel ferrites. The inset compares the Co K-edge RXS of the (0,0,2) and (0,0,6) reflections in $CoFe_2O_4$.

These results confirm that the pre-peak resonance originates from *dq* transitions at the tetrahedral Fe atoms and the main-edge resonance is due to the anisotropy of the trigonal (   ) point symmetry of octahedral B sites of the spinel structure. The azimuthal dependence corroborates the *dd* character of the main-edge resonance, which does not depend on either the type or the formal valency of the transition-metal atom that occupies the octahedral site. We note here that we have observed anisotropic RXS, mainly of a structural origin, in a system where the local atom environment is nearly isotropic since there is not a significant distortion of the ligand configuration [43].

(3) Recently, the correlation of OO and magnetic ordering has been studied on layered cobalt oxides, $RBaCo_2O_{5.5}$ with R=rare-earth [39,40]. Resonances have been observed at the Co K edge for (0,*k*,0) reflections with *k* odd in $TbBaCo_2O_{5.5}$ whereas (*h*,0,0) reflections with *h* odd were not detected either on or off resonance in any phase transition [39]. The cusp of the resonant scattering is either up –(0,3,0) and (0,7,0)- or down –(0,1,0) and (0,5,0)- depending on the *k* value. This behavior arises from displacements of the Tb and Ba ions from the ideal tetragonal positions while the occurrence of RXS and its azimuthal dependence comes form the presence of two different environments for Co atoms (octhedron and pyramid), ordered along the *b* axis. Therefore, resonant (0,*k*,0) reflections with *k* odd corresponds to ATS reflections and no further contribution due to OO has been deduced from this experiment [40]. On the other hand, superlattice (1,0,*l*) with *l* even reflections were reported occurring at the metal-insulator transition in $GdBaCo_2O_{5.5-x}$ [40]. RXS was observed at these reflections, which is sensitive to the difference in anisotropy of adjacent Co pyramids and/or Co octahedral and it has been associated to the presence of OO.

## 5. Conclusions

RXS has demonstrated its potential to solve old and new problems on strong correlated transition-metal oxides. The results on the low temperature insulating phase of magnetite has re-opened the discussion on the origin of the Verwey transition. The classical CO model has been overcome and many recent theoretical papers gives a more realistic interpretation without invoking to ionic ordering. In general, all these RXS results discard the description of mixed-valence transition-metal oxides as a bimodal distribution of integer valence states and the so-called CO phase implies the segregation in different crsytallographic sites whose electronic differences are mainly determined by the structural distortions. These structural distortions induces a different charge density on different crystallographic sites and in some cases, an electronic anisotropy by lowering the local symmetry. Finally, RXS has been also fundamental to determine the type of ordering in many cases, such as $Bi_{0.67}Sr_{0.33}MnO_3$ and $La_{0.33}Sr_{0.67}FeO_3$.

**Acknowledgement**
The authors thank financial support from the Spanish MICINN (FIS08-03951 project) and DGA (Camrads). We also acknowledge ESRF for granting beam time and technical support.

**References**
[1] Materlik G, Spark C J and Fisher K 1994 *Resonant Anomalous X-ray Scattering, Theory and Applications* (Amsterdam: North-Holland/Elsevier Science B. V.)
[2] Stragier H, Cross J O, Rehr J J, Sorensen L B, Bouldin C E and Woicik J C 1992 *Phys. Rev. Lett.* **21** 3064
[3] Proietti M G, Renevier H, Hodeau J L, García J, Bérar J F and Wolfers P 1999 *Phys. Rev. B*. **59** 5479
[4] Templeton D H and Templeton L K 1980 *Acta Cryst. A* **36** 237
[5] Blume M 1994 *Resonant Anomalous X-ray Scattering, Theory and Applications* ed Materlik G, Spark C J and Fisher K (Amsterdam: North-Holland/Elsevier Science B. V.) pp 495-515
[6] Hannon J P, Trammel G T, Blume M and Gibbs D 1988 *Phys. Rev. Lett.* **61** 1245
[7] Di Matteo S, Joly Y and Natoli C R 2005 *Phys. Rev. B* **72** 144406
[8] Verwey E J W 1939 *Nature* **144** 327
[9] Hagiwara K, Kanazawa M, Horie K, Kokubun J and Ishida K 1999 *J. Phys. Soc. Jpn.* **68** 1592
[10] García J, Subías G, Proietti M G, Renevier H, Joly Y, Hodeau J L, Blasco J, Sánchez M C and Bérar J F 2000 *Phys. Rev. Lett.* **85** 578
[11] García J, Subías G, Proietti M G, Blasco J, Renevier H, Hodeau J L and Joly Y 2001 *Phys. Rev. B* **63** 054110
[12] Izimu M, Koetzle T F, Shirane G, Chikazumi S, Matsui M and Todo S 1982 *Acta Cryst. B* **38** 2121
[13] Zuo J M, Spence C H and Petuskey W 1990 *Phys. Rev. B* **42** 8451
[14] Wright J P, Attfield J P and Radaelli P G 2001 *Phys. Rev. Lett.* **87** 266401; 2002 *Phys. Rev. B* **66** 214422
[15] García J, Subías G, Herrero-Martín J, Blasco J, Cuartero V, Sánchez M C, Mazzoli C and Yakhou F 2009 *Phys. Rev. Lett.* **102** 176405
[16] Subías G, García J, Blasco J, Proietti M G, Renevier H and Sánchez M C 2004 *Phys. Rev. Lett.* **93** 156408
[17] Nazarenko E, Lorenzo J E, Joly Y, Hodeau J L, Mannix D and Marin C 2006 *Phys. Rev. Lett.* **97** 056403
[18] Joly J, Lorenzo J E, Nazarenko E, Hodeau J L, Mannix D and Marin C 2008 *Phys. Rev.B* **78** 134110
[19] Wilkins S B, Di Matteo S, Beale T A W, Joly Y, Mazzoli C, Hatton P D, Bencok P, Yakhou F and Brabers V A M 2009 *Phys. Rev. B* **79** 201102(R)
[20] Radaelli P G, Cox D E, Marezio M and Cheong S W 1997 *Phys. Rev. B* **55** 3015
[21] Herrero-Martín J, García J, Subías G, Blasco J and Sánchez M C 2004 *Phys. Rev. B* **70** 024408

[22] Subías G, García J, Beran P, Nevriva M, Sánchez M C and García-Muñoz J L 2006 *Phys. Rev. B* **73** 205107
[23] García J, Sánchez M C, Blasco J, Subías G and Proietti M G 2001 *J. Phys.: Condens. Matter* **13** 3243
[24] Subías G, Sánchez M C, García J, Blasco J, Herrero-Martín J, Mazzoli C, Beran P, Nevriva M, and García-Muñoz J L 2008 *J. Phys.: Condens. Matter* **20** 235211
[25] Park S K, Ishikawa T, Tokura Y, Li J Q and Matsui Y 1999 *Phys. Rev. B* **60** 10788
[26] Herrero-Martín J, Subías G, García J, Blasco J and Sánchez M C 2009 *Phys. Rev. B* **79** 045121
[27] Blasco J, Aznar B, García J, Subías G, Herrero-Martín J and Stankiewicz J 2008 *Phys. Rev. B* **77** 054107
[28] Dmitrienko V E 1983 *Acta Cryst. A* **39** 29
[29] Dmitrienko V E 1984 *Acta Cryst. A* **40** 89
[30] Murakami Y, Kawada H, Kawata H, Tanaka M, Arima T, Moritomo Y and Tokura Y 1998 *Phys. Rev. Lett.* **80** 1932
[31] Benfatto M, Joly Y and Natoli C R 1999 *Phys. Rev. Lett.* **83** 636
[32] Murakami Y, Hill J P, Gibbs D, Blume M, Koyama I, Tanaka M, Kawata H, Arima T, Tokura Y, Hirota K and Endoh Y 1998 *Phys. Rev. Lett.* **81** 582
[33] Zimmermann M, Hill J P, Gibbs D, Blume M, Casa D, Keimer B, Murakami Y, Tomioka Y and Tokura Y 1999 *Phys. Rev. Lett.* **83** 4872
[34] Endoh Y, Hirota K, Ishihara S, Okamoto S, Murakami Y, Nishizawa A, Fukuda T, Kimura H, Nojiri H, Kaneko K and Maekawa S 1999 *Phys. Rev. Lett.* **82** 4328
[35] Di Matteo S, Chatterji T, Joly Y, Stunault A, Paixao J A, Suryanarayanan R, Dhalenne G and Revcolevschi A 2003 *Phys. Rev. B* **68** 024414
[36] Subías G, Herrero-Martín J, García J, Blasco J, Mazzoli C, Hatada K, Di Matteo S and Natoli C R 2007 *Phys. Rev. B* **75** 235101
[37] Paolasini L, Caciuffo R, Sollier A, Ghigna P and Altarelli M 2002 *Phys. Rev. Lett.* **88** 106403
[38] Subías G, García J, Proietti M G, Blasco J, Renevier H, Hodeau J L and Sánchez M C 2004 *Phys. Rev. B* **70** 2155105
[39] Blasco J, García J, Subías G, Renevier H, Stingaciu M, Conder K and Herrero-Martín J 2008 *Phys. Rev. B* **78** 054123
[40] García-Fernández M, Scagnoli V, Staub U, Mulders A M, Janousch M, Bodenthin Y, Meister D, Patterson B D, Mirone A, Tanaka Y, Nakamura T, Grenier S, Huang Y and Conder K 2008 *Phys. Rev. B* **78** 054424
[41] Goodenough J B 1963 *Magnetism and Chemical Bond* (New York:Interscience)
[42] Benfatto M, Della Longa S and Natoli C R 2003 *J. Synchrotron Radiat.* **10** 51
[43] Subías G, García J and Blasco J 2005 *Phys. Rev. B* **71** 155103